\documentclass[prl,amsmath,amssymb,twocolumn,superscriptaddress]{revtex4}
\usepackage{amsmath,amssymb,graphicx}
\usepackage{dcolumn}

\begin{document}


\title{Subdiffusive Energy Transport and Antipersistent Correlations Due to the Scattering of Phonons and Discrete Breathers}

\author{Daxing Xiong}
\email{xmuxdx@163.com}
\affiliation{MinJiang Collaborative Center for Theoretical Physics, College of Physics and Electronic Information Engineering, Minjiang University, Fuzhou 350108, China}

\author{Jianjin Wang}
\email{phywjj@qq.com}
\affiliation{Department of Physics, Jiangxi Science and Technology Normal University, Nanchang 330013, Jiangxi, China}

\begin{abstract}
While there are many physical processes showing subdiffusion and some useful particle models for understanding the underlying mechanisms have been established, a systematic study of subdiffusive energy transport is still lacking. Here we present convincing evidence that the energy subdiffusion and its antipersistent correlations take place in a Hamiltonian lattice system with both harmonic nearest-neighbor and anharmonic long-range interactions. We further understand the underlying mechanisms from the scattering of phonons and discrete breathers. Our result sheds new light on understanding the extremely slow energy transport.
\end{abstract}

\maketitle

Subdiffusive motion differs from normal diffusion in that the mean squared displacement (MSD) does not grow linearly in time $t$ but scales as $t^{\beta}$ with $\beta <1$~\cite{PhysRP2015}. This type of significantly slow diffusion motion has been observed in many diverse complex systems, ranging from amorphous semiconductors~\cite{Sub-1} to geology~\cite{Sub-2}, from weak turbulence in liquid crystals~\cite{Sub-3} to particle motion in living cells~\cite{Sub-4,Sub-5,Sub-6}, and from suspension of colloidal beads~\cite{Sub-7} to quantum and spin systems~\cite{Sub-8,Sub-9}. To understand the underlying mechanisms, several prominent physical models such as the confined continuous time random walk and the fractional Brownian motion have been proposed~\cite{SubRep-1,SubRep-2,SubRep-3}. However, most of these models are devoted to the motion of particles, the counterparts of energy transport of subdiffusion type which are ubiquitous as well are less studied. What are the underlying mechanisms of energy subdiffusion and how do the features of energy subdiffusion emerge when dealing with a many-body Hamiltonian system are still challenging.

Due to the finiteness of the Poincar\'{e} recurrence time, it is argued that the subdiffusive motion cannot be achieved in conventional Hamiltonian systems in the absence of disorder~\cite{Book2007}. Therefore up to now only quite few examples of subdiffusive energy transport in Hamiltonian systems have been reported. The first convincing one is the thermal transport in a billiard channel model in a very special configuration, which shows $\beta \simeq 0.86$~\cite{Alonson2002}. However, in billiards the heat transport is performed by the diffusing particles. Thus the billiard is essentially a particle model~\cite{Denisov2003}, in contrast to the many-body lattice Hamiltonian systems where the energy transport is conducted by phonons, i.e., the collective excitation modes or quasi-particles. This raises an interesting question if energy subdiffusion can exist in lattices~\cite{WangJ2003}. Recently, this question was partially answered by studying subdiffusive energy transport in a special Hamiltonian mean-field (HMF) model~\cite{Sube-1,Sube-2}, where some of the signatures of subdiffusion like the vanishing of heat current have been revealed. However, other significant features such as the antipersistent~\cite{Antipersistent} correlations that were frequently observed in the subdiffusion of particles have not yet been explored. Furthermore, what is the origin of this antipersistence is still open.

In this Letter we show that the features of energy subdiffusion, such as the subdiffusive scaling of the probability distribution function (PDF), the sublinear increase of MSD in $t$, and the antipersistent heat current autocorrelation can emerge in a many-body Hamiltonian lattice system with both harmonic nearest-neighbor (NN) and anharmonic long-range interactions (LRIs). The NN harmonic interaction enables the excitation of phonons, while the anharmonic LRIs, with appropriate strengths, favor the excitation of discrete breathers (DBs). These DBs can be the scatters of phonons. Therefore various scattering processes invoke rich transports and present different antipersistence for energy subdiffusion. In this sense our result unveils new underlying physical pictures for subdiffusive motion.

We consider a one-dimensional Hamiltonian lattice of $N$ particles under the Born-von Karman periodic boundary conditions~\cite{BornvonKarman} represented by
\begin{equation}
H=\sum_{i=1}^N \left[\frac{p_{i}^{2}}{2}+ \frac{1}{2} (x_{i+1}-x_i)^2 + \frac{1}{4} \sum_{r=1}^{\tilde{N}} (x_{i+r}-x_i)^4  \right].
\label{HH}
\end{equation}
Here, $x_{i}$ and $p_{i}$ are two canonically conjugated variables with $i$ the index of the particle; all other relevant quantities like the particle's mass and the lattice constant are dimensionless and set to be unity. The interparticle interactions include two separate terms. The first one denotes the harmonic NN coupling and the second one is the anharmonic LRIs~\cite{Note2}. The periodic boundary conditions make our system like a ring.

For the LRIs we first consider the system in the presence of global interactions, i.e., $\tilde{N}=\frac{N}{2}-1$. This setup is thus similar to that of the HMF model. This latter type of LRIs, together with the harmonic NN interaction, thus favor the excitation of both phonons and DBs. However, the HMF model with both NN and LRIs always supports superdiffusive transport~\cite{Sube-2}. Fortunately, in the following we are able to reveal the subdiffusive motion.

To capture the PDF of energy diffusion in such a ring at a given temperature ($T=0.5$ is considered), we employ the equilibrium spatiotemporal correlation function (see Refs.~\cite{Zhao2006,Zhao2013,Xiong2020} for details)
\begin{equation}
\rho_{E}(m,t)=\frac{\langle \Delta E_{i+m}(t) \Delta E_{i}(0) \rangle}{\langle \Delta E_{i}(0) \Delta E_{i}(0) \rangle}
\end{equation}
of the local energy $E_i=\frac{p_{i}^{2}}{2}+ \frac{1}{2} (x_{i+1}-x_i)^2+ \frac{1}{4} \sum_{r =1}^{\tilde{N}} (x_{i+r}-x_i)^4$ for a canonical system. Here, due to the translational invariance, the correlation depends only on the relative distance $m$; $\langle \cdot \rangle$ represents the spatiotemporal average; $\Delta E_{i}(t)=E_{i}(t)-\langle E \rangle$.

To calculate $\rho_{E}(m,t)$, we set the total system size $N=4096$, which ensures that an initial energy fluctuation located at the center can spread out at a time at least up
to $t=2000$. We use the velocity-Verlet algorithm~\cite{Verlet} with a small time step $ 0.01$ to evolve the system. We adopt a Fast Fourier Transform (FFT)~\cite{FFT} algorithm to accelerate our computations. We utilize an ensemble of size about $8 \times 10^9$.
\begin{figure}[!t]
\vskip-0.2cm
\includegraphics[width=8.8cm]{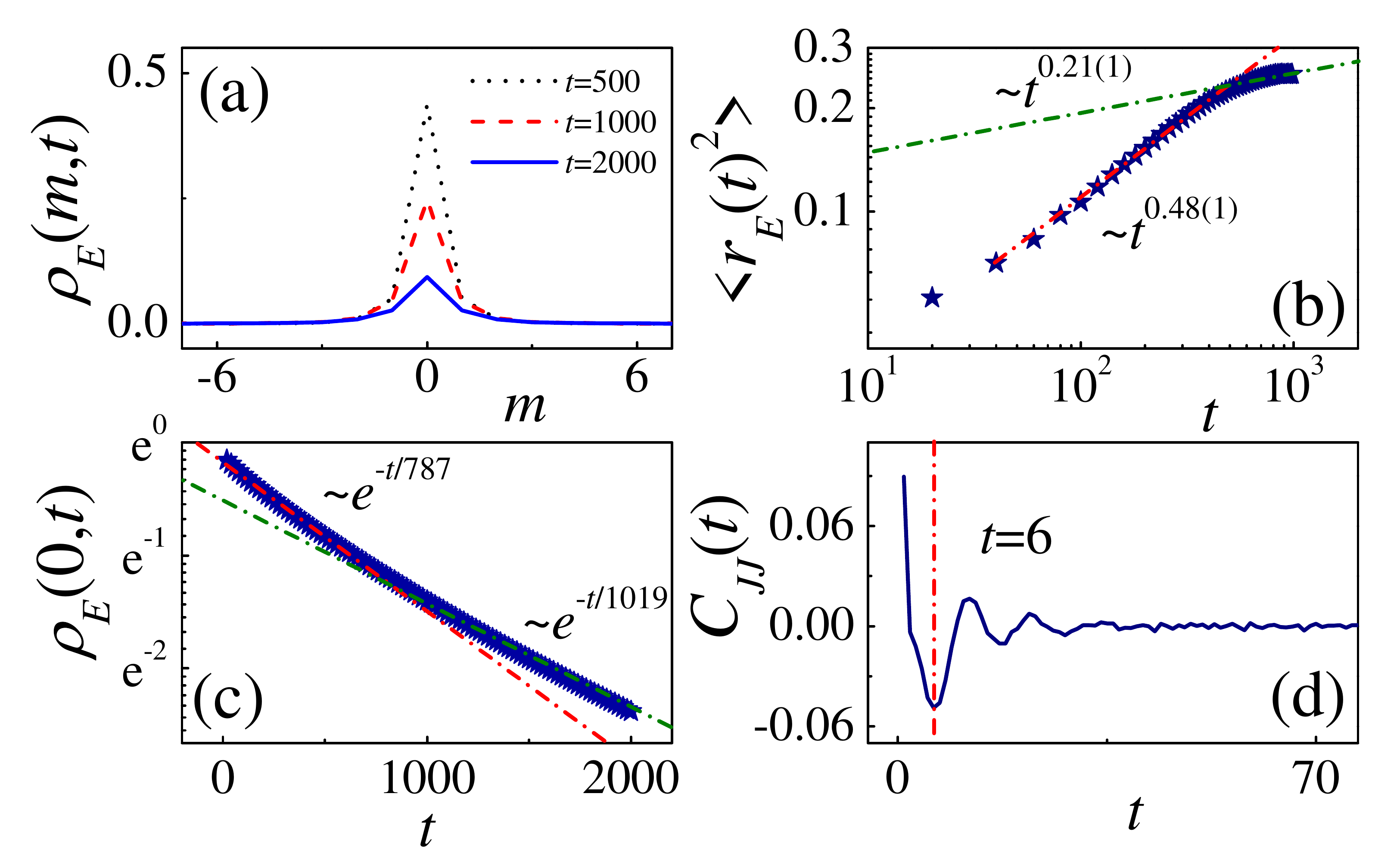}
\vskip-0.3cm
\caption{New type of subdiffusive energy transport in the system with $\tilde{N}=\frac{N}{2}-1$:  (a) $\rho_{E}(m,t)$ for several times $t$; (b) $\langle r_E(t)^2 \rangle$ versus $t$ showing double sublinear scalings with $\beta_1 \simeq 0.48$ and $\beta_2 \simeq 0.21$; (c) $\rho_E(0,t)$ versus $t$ showing a two-stage exponential decay with characteristic times $\tau_1 \simeq 787$ and $\tau_2 \simeq 1019$; (d) the equilibrium heat current auto-correlation $C_{JJ}(t)$ vs $t$ indicating the antipersistent correlations, where the first sharp negative minimum is at $t=6$.} \label{fig:1}
\end{figure}

Figure~\ref{fig:1}(a) depicts $\rho_E(m,t)$ for several long times showing an extremely slow spread of the local energy fluctuations $\Delta E_{i}(t)$, i.e., the PDFs are quite localized and only cover few lattice sites. This gives a first visual sign of subdiffusive energy transport. To further check this, Fig.~\ref{fig:1}(b) depicts the MSD $\langle r_E(t)^2 \rangle$ calculated by
\begin{equation}
\langle r_E(t)^2 \rangle=\sum_{m=-N/2}^{N/2} m^2 \rho_E(m,t)
\end{equation}
of $\rho_E(m,t)$. $\langle r_E(t)^2 \rangle \sim t^{0.48}$ for a short $t$ and $\langle r_E(t)^2 \rangle \sim t^{0.21}$ at long times are numerically fitted~\cite{Note}. Interestingly, it not only confirms our above conjecture, but also suggests new insight, i.e., double sublinear scalings of MSD at short and long times. With this finding in mind, we next check the time dependence of $\rho_E(0,t)$ in Fig.~\ref{fig:1}(c). In the framework of random walk model, $\rho_E(0,t)$ means the probability of the energy being at initial position at some $t$. For normal and super energy diffusion, one expects $\rho_E(0,t) \sim t^{-\beta}$ with $\beta=\frac{1}{2}$ and $\beta > \frac{1}{2}$, respectively~\cite{PhysRP2015,Levywalk}. But for the subdiffusive regime observed here, remarkably our best fitting no longer follows a single scaling of $\rho_E(0,t) \sim t^{-\beta}$, instead it shows a manner of two-stage exponential decay $\rho_E(0,t) \sim \exp (-\frac{t}{\tau})$ with two characteristic times: $\tau_1 \simeq 787$ for short times and $\tau_2 \simeq 1019$ at long times. These two characteristic times ($\tau_1 <\tau_2$) seems to be consistent to the judgement of double sublinear scalings ($\beta_1 > \beta_2$) of MSD.

The feature of subdiffusion can also been revealed by studying the system's equilibrium heat current autocorrelation
\begin{equation}
C_{JJ}(t)=\langle J_{\rm tot} (t) J_{\rm tot}(0) \rangle,
\end{equation}
where $J_{\rm tot}=\sum_{i=1}^{N} p_i \left[(x_{i+1}-x_i)+ \sum_{r=1}^{\tilde{N}} (x_{i+r}-x_i)^3\right]$ is the total heat current along the lattice. $C_{JJ}(t)$ is related to the thermal conductivity $\kappa$ based on the Green-Kubo formula $\kappa=\lim_{\tau \rightarrow \infty} \lim_{N \rightarrow \infty} \frac{1}{k_B N T^2} \int_{0}^{\tau} C_{JJ}(t) dt$. For diffusive and superdiffusive transport, one would expect $C_{JJ} \sim \exp{(-\frac{t}{\nu})}$ and $C_{JJ}(t) \sim t^{-\gamma}$, respectively. So $\kappa$ is finite (divergent) for diffusive (superdiffusive) transport. For subdiffusive transport, however, $\kappa$ should be vanishing in the thermodynamic limit, indicating the system to be a thermal insulator~\cite{Denisov2003,WangJ2003,Sube-1,Sube-2}. To support this, mathematically $C_{JJ}(t)$ has to change sign at least once in order to cause the Green-Kubo integral vanishing since it is a continuous function of $t$. Physically, particularly in the framework of particle models for subdiffusion~\cite{SubRep-1,SubRep-2,SubRep-3}, this means the property of \emph{antipersistence}, i.e., a period of energy transport with positive heat current is typically followed by a period of transport with negative current. Our plot of the antipersistent heat current autocorrelations in Fig.~\ref{fig:1}(d) again supports the subdiffusive motion. These correlations show a first sharp negative minimum at $t=6$ and a few local negative minima following behind. This is similar to the antipersistent velocity autocorrelations in the particle subdiffusion~\cite{PhysRP2015,Sub-4,Sub-5}, but we would like to stress that to our best knowledge, this is for the first time to attract our attention in the energy transport.

Combining the above evidence, we suggest a phonon-DBs scattering picture to understand this new type of energy subdiffusion (we have made a vivid animation for phonon-DBs scattering in zero-temperature systems, see movies in Supplementary Material (SM)~\cite{SM}). Indeed, phonons, as the main heat carriers, are originated from the harmonic NN coupling in the Hamiltonian~\eqref{HH}. DBs, supposed as the viscoelastic elements~\cite{Sub-5}, come from the anharmonic LRIs~\cite{Sube-2}. As the phonon waves move through DBs, which is regarded as a scatter, phonons will be partially reflected, partially pass through, and accompanied with energy loss ~\cite{Scattering1,Scattering2}. The antipersistent correlations shown in Fig.~\ref{fig:1}(d) implies that the reflection takes place for short times and that DBs are finally in the majority. The double sublinear scalings ($\beta_2 < \beta_1$) of MSD [Fig.~\ref{fig:1}(b)] and double characteristic times ($\tau_2 > \tau_1$) of $\rho_E(0,t)$ [Fig.~\ref{fig:1}(c)] thus suggests that statistically, only a few rounds of the phonon-DBs scattering processes play a primary role. Therefore we infer that the first-stage energy relaxation is induced by phonon-DBs scattering while the second-stage relatively slow relaxation is mainly caused by DBs.
\begin{figure}[!t]
\includegraphics[width=9cm]{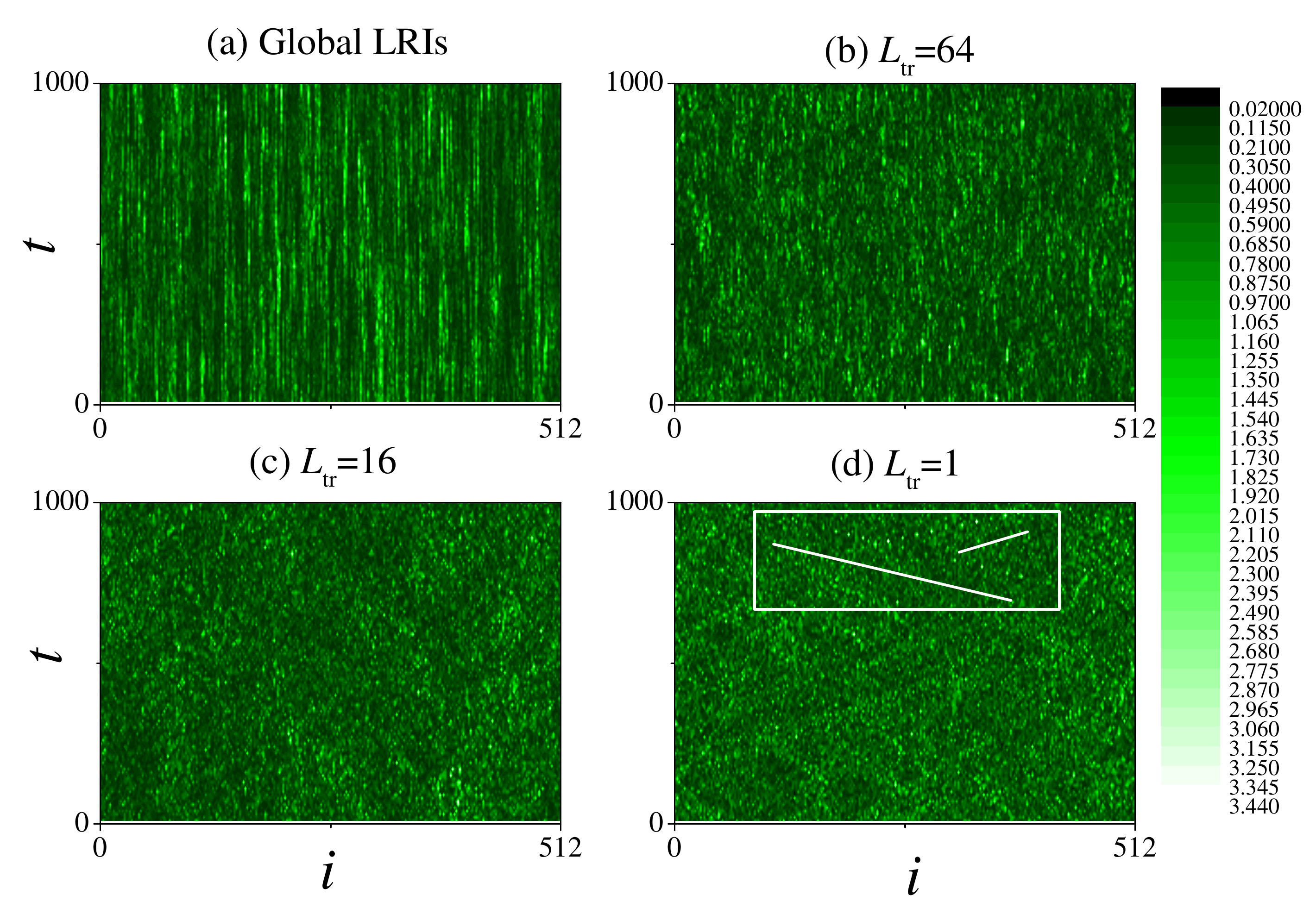}
\vskip-0.2cm
\caption{DBs in finite-temperatures: Spatiotemporal evolution of energy densities $E_i(t)$ for the system ($N=512$) at thermal equilibrium for a time scale $t=1000$ with a time step $\Delta t=10$ : (a) the global LRIs ($\tilde{N}=\frac{N}{2}-1$); (b)-(d) the truncated LRIs with $L_{\rm tr}=64$, $L_{\rm tr}=16$, and $L_{\rm tr}=1$ (the NN coupling case), respectively. Those marked regimes in (d) denote the mobile excitations.} \label{fig:2}
\end{figure}

To further verify the phonon-DBs scattering picture in finite-temperatures, we next consider a system with the truncated LRIs. This means that we set $\tilde{N}=L_{\rm tr}<\frac{N}{2}-1$. This setup will weaken the strength of DBs and make the phonon-DBs scattering more evident. This is indeed indicated in the movies in zero-temperature systems (see SM~\cite{SM}). However, to show DBs in finite-temperatures is more challenging and we study the spatiotemporal evolutions of local energy densities $E_i(t)$ under the equilibrium state to achieve this~\cite{Xiong2020}. To do this, the ring is first thermalized to $T=0.5$, then the thermal baths are removed and the results are recorded and displayed for a time scale $t=1000$ by a suitable time step $\Delta t=10$. To avoid huge data to display, we now consider a short ring of $N=512$ instead. Fig.~\ref{fig:2} depicts the spatiotemporal evolutions of the systems with  $\tilde{N}= \frac{N}{2}-1$ and several $L_{\rm tr}$. As indicated, DBs represented by localizations for $\tilde{N}= \frac{N}{2}-1$ are obviously stronger than those in the truncated LRIs [see~Fig.~\ref{fig:2}(a,b)]. In addition, with the decrease of $L_{\rm tr}$, the phonon-DBs scattering leads DBs less evident and thus the spatiotemporal evolutions seem to show a mixed picture~[see~Fig.~\ref{fig:2}(c)]. Finally, if $L_{\rm tr}$ is further decreased to $L_{\rm tr}=1$, i.e., a system with NN couplings only, one can clearly identify the signatures of moving excitations~[see~Fig.~\ref{fig:2}(d)]. These excitations correspond to the long wavelength phonons or solitary waves, which are usually regarded as the microscopic origin of the superdiffusive thermal transport in the short-range interacting systems. So we here further confirm that even in the finite-temperature systems it is the competition between the harmonic NN couplings and the anharmonic LRIs which gives the scattering of phonons and DBs.
\begin{figure}[!t]
\includegraphics[width=8.8cm]{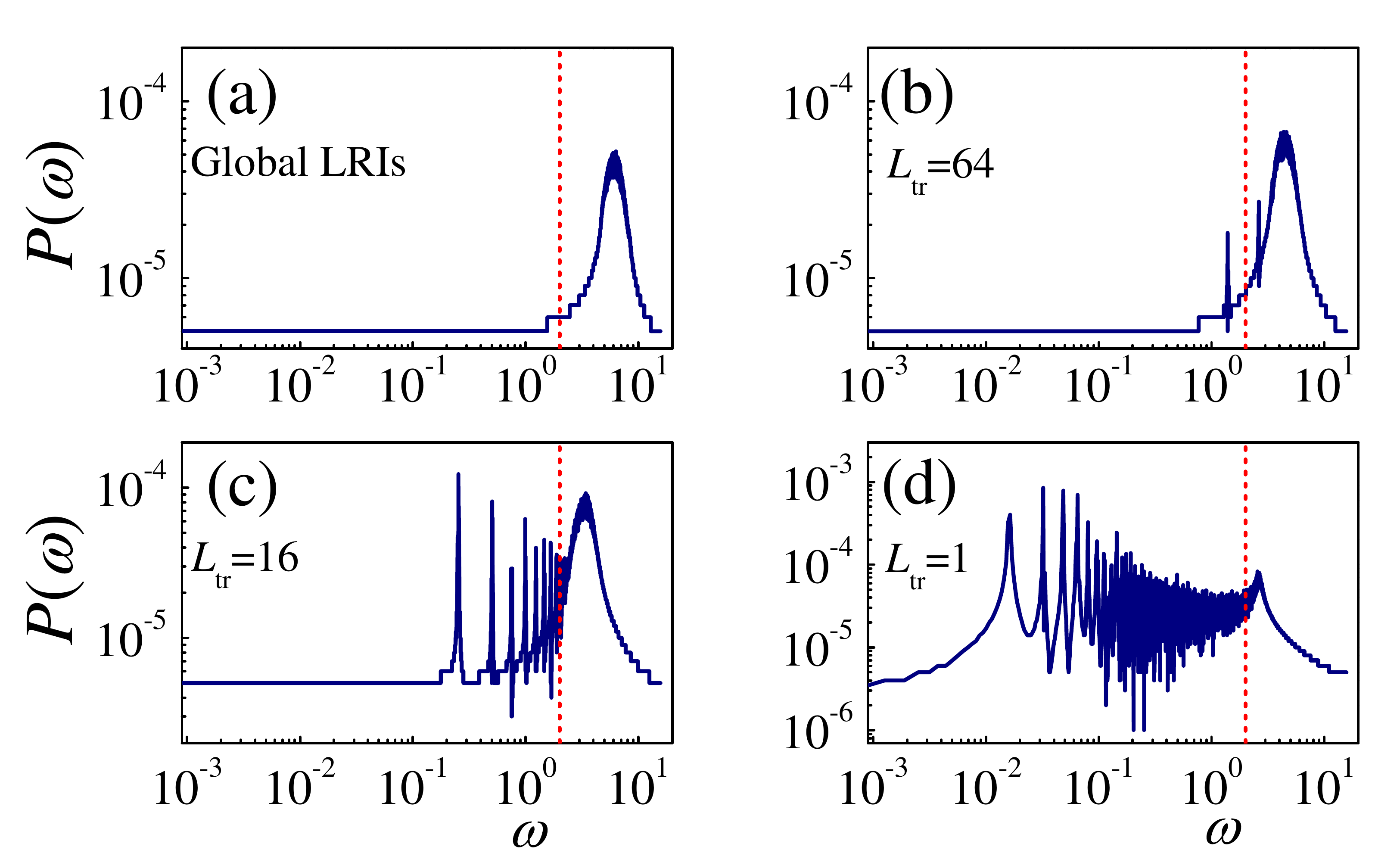}
\vskip-0.2cm
\caption{Phonon-DBs scattering in finite-temperatures: The spectra $P(\omega)$ of the thermal fluctuations ($N=512$): (a) the global LRIs ($\tilde{N}=\frac{N}{2}-1$); (b)-(d) the truncated LRIs with $L_{\rm tr}=64$, $L_{\rm tr}=16$, and $L_{\rm tr}=1$ (the NN coupling case), respectively. The vertical dashed lines indicate the Brillouin zone $\omega=2$ of linear phonons.} \label{fig:3}
\end{figure}

More evidence of phonon-DBs scattering in finite-temperatures can be revealed by studying the typical spectrum $P(\omega)$ of the thermal fluctuations. This is achieved by performing a frequency $\omega$ analysis of the equilibrium momentum $p(t)$ along the ring: $P(\omega)=\lim_{\tau \rightarrow \infty} \frac{1}{\tau}
\int_{0}^{\tau} p (t) \exp (- j \mit{\omega} t) \rm{d}
\mit{t}$~\cite{NianbeiLi_Review}, where $j^2=-1$. Fig.~\ref{fig:3} depicts $P(\omega)$ for the same setups as Fig.~\ref{fig:2}. To distinguish the long wavelength phonon waves, we adopt a log-log plot to show the low frequency components clearly. Firstly, for a large enough $L_{\rm tr}$, most of the frequency components are lying in above the Brillouin zone $\omega=2$ of linear phonons confirming the properties of DBs [see~Fig.~\ref{fig:3}(a,b)]. Secondly, it is known that in lattices with only NN couplings supporting superdiffusive transport, the low frequency phonons are weakly damped due to the conserved feature of momentum~\cite{Weakdamp,Xiong2016}. As the NN anharmonicity increases we always see the damping of phonons starting from the high frequency components. This is indeed indicated in Fig.~\ref{fig:3}(d) where the low frequency can be clearly identified (see also~\cite{Xiong2016}). But after introducing the anharmonic LRIs, the situation reverses, i.e., as the anharmonicity increases we always see the attenuation of phonons beginning at the low frequency components [see~Fig.~\ref{fig:3}(b,c)], and finally phonons are immersed in the environment of DBs [see~Fig.~\ref{fig:3}(a)]. This latter property seems to demonstrate the peculiarity of energy subdiffusion in lattices. It again supports the phonon-DBs scattering picture.

\begin{figure}[!t]
\includegraphics[width=8.8cm]{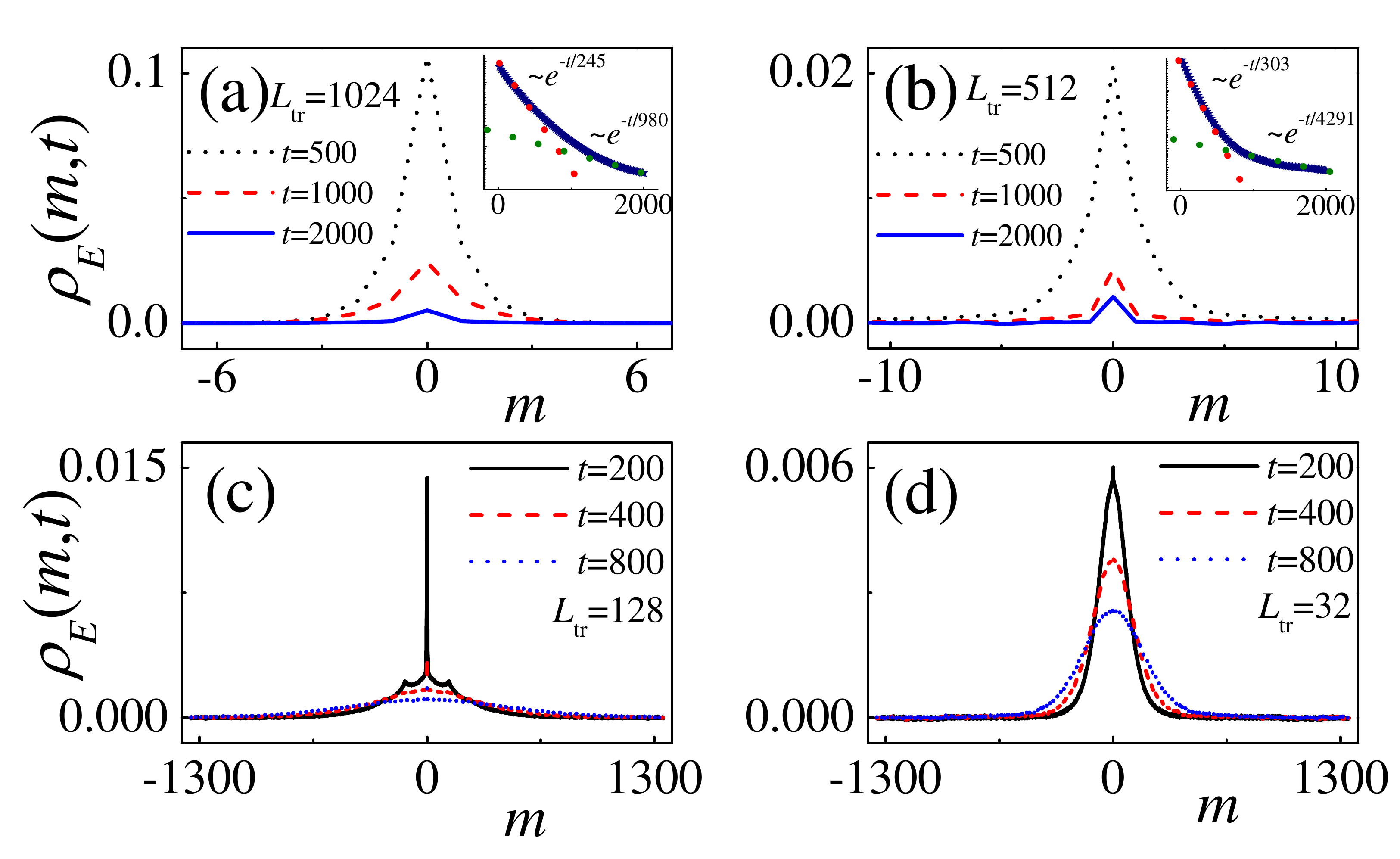}
\vskip-0.2cm
\caption{From subdiffusion to superdiffusion: $\rho_{E}(m,t)$ for several times $t$ for the systems with the truncated LRIs: (a) $L_{\rm tr}=1024$; (b) $L_{\rm tr}=512$; (c) $L_{\rm tr}=128$, and (d) $L_{\rm tr}=32$, respectively ($N=4096$). The insets in (a) and (b) show $\rho_E(0,t)$ vs $t$. Note that the scale of the x-axis is different for each panel.} \label{fig:4}
\end{figure}

We now capture the two-stage energy relaxation and the antipersistent correlations in more detail. Toward that end, we return to the results of $\rho_E(m,t)$ and $C_{JJ}(t)$ but for the systems with the truncated LRIs instead. From Fig.~\ref{fig:4} one expects that for a relatively short $L_{\rm tr}$, phonons will succeed over DBs after the scattering, so energy subdiffusion represented by $\rho_E(m,t)$ would not be seen~[see Figs.~\ref{fig:4}(c,d)]. While for a relative large $L_{\rm tr}$, DBs are in the majority after the scattering, the subdiffusion can thus take place [see Figs.~\ref{fig:4}(a,b)]. Interestingly, for all the subdiffusion observed here, $\rho_E(0,t)$ always shows a two-stage relaxation [see the insets of Figs.~\ref{fig:4}(a,b)]. Therefore whether DBs finally survive most after the scattering is crucial to the emergence of subdiffusive energy transport, and if yes the double-scaled subdiffusive motions seem always to happen.

\begin{figure}[!t]
\includegraphics[width=8.8cm]{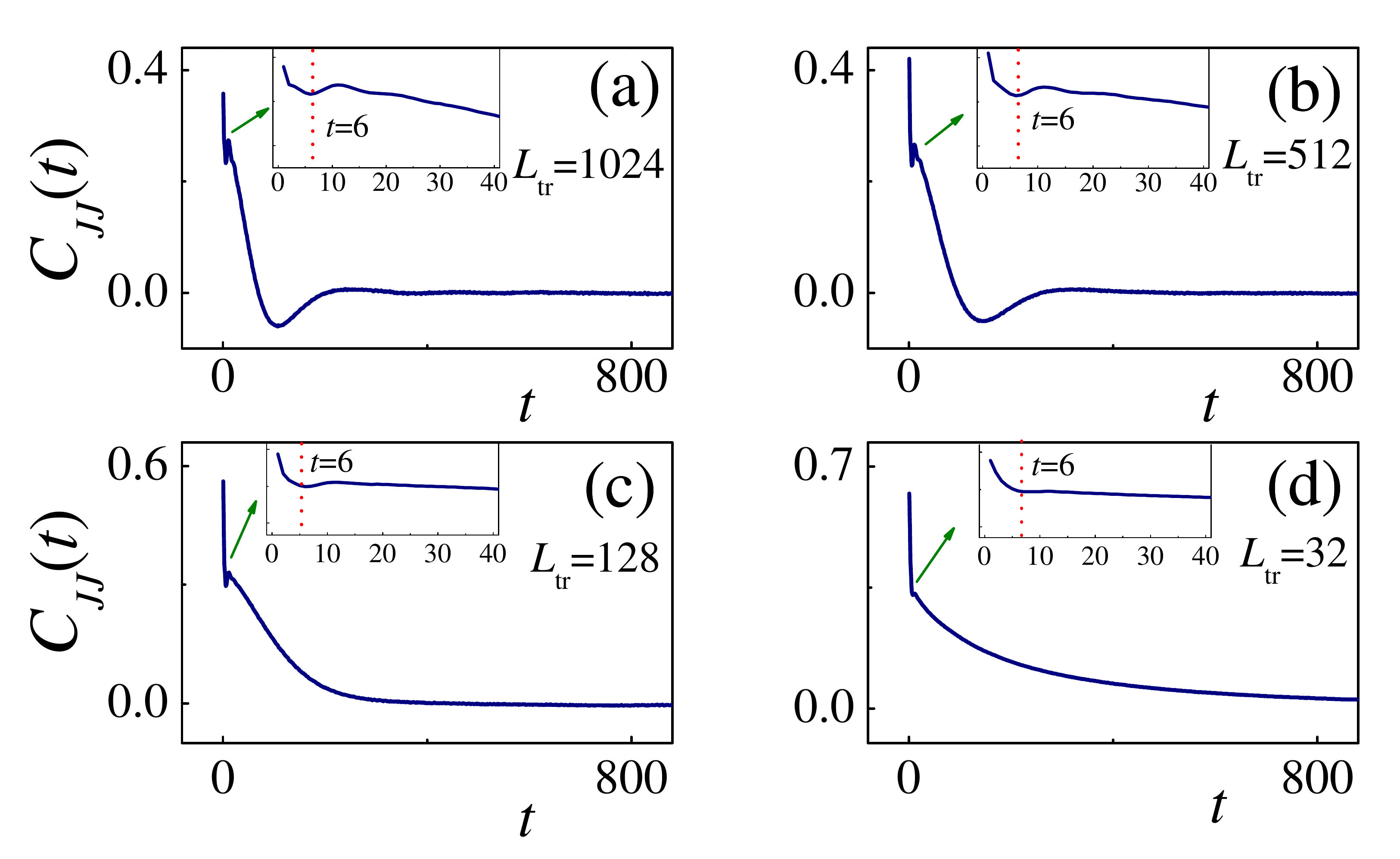}
\vskip-0.2cm
\caption{The details of antipersistence: The equilibrium heat current auto-correlation $C_{JJ}(t)$ vs $t$ for the systems with the truncated LRIs: (a) $L_{\rm tr}=1024$; (b) $L_{\rm tr}=512$; (c) $L_{\rm tr}=128$, and (d) $L_{\rm tr}=32$, respectively ($N=4096$). The insets are a zoom for identifying the first turning point $t=6$ for denoting the antipersistence.} \label{fig:5}
\end{figure}

The antipersistent correlations are still shown in the systems with the truncated LRIs [see~Fig.~\ref{fig:5}]. If $L_{\rm tr} \geq 32$, this antipersistence will always present in our system at a universal short time $t=6$. But compared with the systems with the global LRIs, it turns from a sharp negative minimum to a positive dip. At present we do not know whether this is an intrinsic property of such kind of systems. However, the universality of $t=6$ shown in the systems must imply its same origin from the phonon-DBs scattering. The only distinction is that some kinds of phonon-DBs scattering is strong leading to the sharp negative minimum~[see~Fig.~\ref{fig:1}(d) and movies in SM~\cite{SM}]; while other kinds of phonon-DBs scattering is relatively weak and the antipersistence only presents as a dip~[see~Figs.~\ref{fig:5}(a-c) and movies in SM~\cite{SM}]. Another important information is that $C_{JJ}(t)$ in Figs.~\ref{fig:5}(a,b) show a second antipersistence with negative minima but these minima are not presented in Figs.~\ref{fig:5}(c,d). Combining the results of Figs.~\ref{fig:4}(c,d), one infers that whether $C_{JJ}(t)$ has negative correlations seems crucial for signaling out the subdiffusive motion. Interestingly, the two dips shown in Figs.~\ref{fig:5}(a,b) also support the few rounds of phonon-DBs scattering pictures.

To summarize, we have presented the first detailed investigation of subdiffusive energy transport in a many-body Hamiltonian lattice. This new type of energy subdiffusion show features that are quite similar to the counterparts of particles, e.g., the most important property of the antipersistence. But whatever it is, it would undoubtedly invoke new insight since the energy considered here also contains the information of many-body interactions. Indeed, the overall underlying physics is now understood from phonon-DBs scattering, i.e., the collective linear and nonlinear excitations, rather than the real particles performing random walks in complicated environments. Due to this, some novel and may be general features, such as the two-stage energy relaxation and a universal dip indicating the antipersistent correlation emerge. Inspired by this we are able to further clarify that the negative velocity/current correlations are crucial to judge subdiffusive motion.

Apart from these insights, to find subdiffusive energy transport in Hamiltonian dynamics in itself has its theoretical significance, which would help us check the validity of KAC lemma~\cite{Book2007}. In addition, the thermal conductivity in such kind of systems will be vanishing in the thermodynamic limit, which implies the system to be a thermal insulator. It would thus be fascinating to expect our theoretical understanding here to invoke potential applications in designing new thermal devices.

\begin{acknowledgments}
We thank helpful discussions from Prof.~Youjin Deng and Prof.~Hai-Jun Zhou. D.X. is supported by NSF (Grant No. 2021J02051) of Fujian Province of China; J.W. is supported by NNSF (Grant No. 12105122) of China.
\end{acknowledgments}



\begin{thebibliography}{100}
\bibitem{Note} We do not present more data for the MSD $\langle r_E(t)^2 \rangle$ at large times because of the huge flucutations of $\rho_{E}(m,t)$ for a large $m$ in finite-temperature calculations.
\end{thebibliography}
\end{document}